\documentclass{INTERSPEECH2023}
\usepackage{boldline,multirow,multicol}
\usepackage{graphicx}
\usepackage{subfig}
\usepackage{pgfplots, mathtools}
\usepackage{xcolor}
\usepackage{cite}

\definecolor{turquoise}{cmyk}{0.65,0,0.1,0.3}
\definecolor{purple}{rgb}{0.65,0,0.65}
\definecolor{dark_green}{rgb}{0, 0.5, 0}
\definecolor{orange}{rgb}{0.8, 0.6, 0.2}
\definecolor{red}{rgb}{0.8, 0.2, 0.2}
\definecolor{darkred}{rgb}{0.6, 0.1, 0.05}
\definecolor{blueish}{rgb}{0.0, 0.3, .6}
\definecolor{light_gray}{rgb}{0.7, 0.7, .7}
\definecolor{pink}{rgb}{1, 0, 1}
\definecolor{greyblue}{rgb}{0.25, 0.25, 1}
\definecolor{dark_gray}{rgb}{0.3, 0.3, 0.3}

\newcommand{\tblue}[1]{{\color{blueish}#1}}
\newcommand{\tred}[1]{{\color{darkred}#1}}
\newcommand{\torange}[1]{{\color{orange}#1}}
\newcommand{\tgray}[1]{{\color{dark_gray}#1}}

\newcommand{\fullname}{Multi-level feature Fusion-based Periodicity Analysis Model}
\newcommand{\shortname}{MF-PAM}

\interspeechcameraready

\title{MF-PAM: Accurate Pitch Estimation \\through Periodicity Analysis and Multi-level Feature Fusion}
\name{Woo-Jin Chung$^1$, Doyeon Kim$^1$ Soo-Whan Chung$^2$ Hong-Goo Kang$^1$}
\address{
  $^1$Dept. of Electrical \& Electronic Engineering, Yonsei University, South Korea\\
  $^2$NAVER Cloud, South Korea 
  }
\email{woojinchung@dsp.yonsei.ac.kr
 ehyeon24@dsp.yonsei.ac.kr, soowhan.chung@navercorp.com, hgkang@yonsei.ac.kr}

\begin{document}

\maketitle

\begin{abstract}
We introduce Multi-level feature Fusion-based Periodicity Analysis Model (MF-PAM), a novel deep learning-based pitch estimation model that accurately estimates pitch trajectory in noisy and reverberant acoustic environments.
Our model leverages the periodic characteristics of audio signals and involves two key steps: extracting pitch periodicity using periodic non-periodic convolution (PNP-Conv) blocks and estimating pitch by aggregating multi-level features using a modified bi-directional feature pyramid network (BiFPN). 
We evaluate our model on speech and music datasets and achieve superior pitch estimation performance compared to state-of-the-art baselines while using fewer model parameters. 
Our model achieves 99.20 \% accuracy in pitch estimation on a clean musical dataset. 
Overall, our proposed model provides a promising solution for accurate pitch estimation in challenging acoustic environments and has potential applications in audio signal processing.

\end{abstract}

\noindent\textbf{Index Terms}: Neural pitch estimation, multi-level fusion

\section{Introduction}

Voice is an intrinsic attribute of humans and depends on the physiological articulatory anatomy of each person.
When producing speech, an intricate combination of organs, from the lungs to the mouth and throughout the vocal tract, collaborates to produce an individual's unique voice.
In particular, pitch or fundamental frequency is widely regarded as a prominent characteristic of a speaker's voice among other acoustic features, and it is essential for various speech-oriented tasks such as speech enhancement~\cite{makhijani2013speech,wang2022hgcn}, speech separation~\cite{han2012classification,wang2019pitch}, speech synthesis~\cite{song2015improved,reddy2020excitation}, and speaker verification~\cite{cheng1998speaker,ghahremani2014pitch}. 

Previously, stochastic approaches such as normalized auto-correlation or zero-crossing intervals were primarily used to estimate the periodicity of speech in the time domain~\cite{de2002yin,morise2009fast,morise2010rapid}.
Other techniques such as difference functions and spectral analysis have been used to explore the harmonicity of signals in the frequency domain~\cite{boersma1993accurate}.
pYIN~\cite{mauch2014pyin} used the local minima of the cumulative mean normalized difference function and hidden Markov models for probabilistic modification, whereas SWIPE~\cite{camacho2008sawtooth} estimated the pitch in the frequency domain using the sawtooth waveform spectrum.
Hybrid approaches that analyzed both the time and frequency domains propose for more stable performance~\cite{wu2003multipitch, kasi2002yet}.
However, despite their lightweight approach, these stochastic methods have shown unstable pitch estimation performances due to various limitations.
Typically, an accurate estimation of pitch is challenging due to its dependence on multiple factors including intonation, emotion, and even physiological factors that may vary over time.
Moreover, pitch estimation in observed speech remains a challenging task due to the potential distortions caused by environmental factors.

Deep learning techniques in speech processing have significantly improved the performance of pitch estimation.
In~\cite{han2014neural}, the authors have proposed effective estimation networks based on the sequential modeling of neural networks.
CREPE~\cite{kim2018crepe} leveraged convolution neural networks~(CNNs) considering the noise distortion, while DeepF0~\cite{singh2021deepf0} utilized a dilated causal convolution network for large receptive field observation.
Both methods have proved that neural networks are effectively used to analyze the acoustic characteristics of speech signals.
More recent studies have focused on the development of models that consider acoustic characteristics rather than relying solely on neural networks.
SPICE~\cite{gfeller2020spice} analyzed the pitch shift mapped by constant-Q transform~(CQT), and HarmoF0~\cite{wei2022harmof0} captured the harmonic structure closely related to pitch from a log-spectrogram using multiple rates dilated causal convolution.
These studies demonstrated that considering the acoustic characteristics of speech signals improves the performance of pitch estimation.

\begin{figure*}[!ht]
    \centering
    \includegraphics[width=0.9\linewidth]{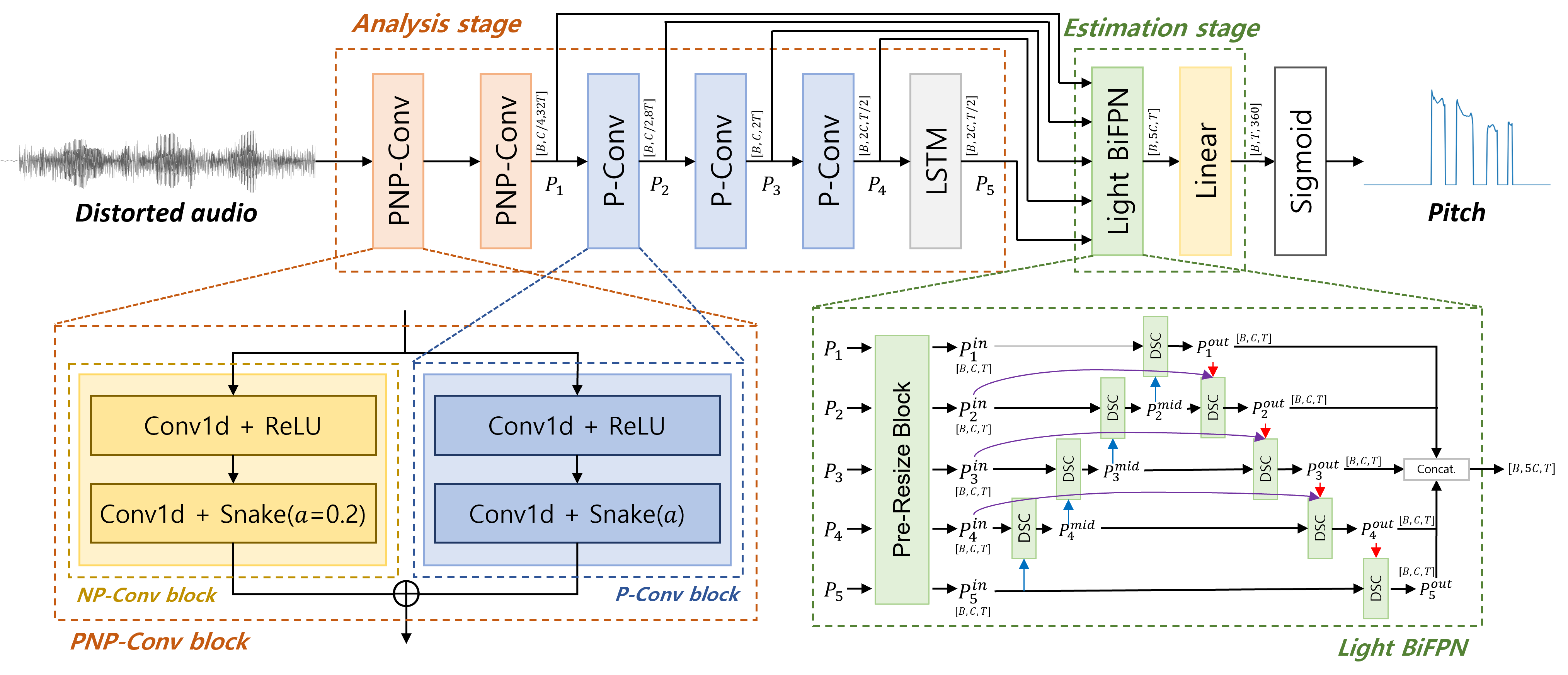}
    \vspace{-5pt}
    \caption{Illustration of the proposed model, MF-PAM.}
    \vspace{-10pt}
    \label{total_fig}
\end{figure*}

In this paper, we propose a novel lightweight pitch estimation model, \fullname~(\shortname), which operates on the raw audio waveform.
The proposed model is composed of two stages: analysis and estimation.
During the analysis stage, \shortname~extracts the periodicity of the feature maps from the input speech utilizing two submodules. 
The low-level submodule distinguishes the periodic and non-periodic characteristics by using periodic and non-periodic convolution (PNP-Conv) blocks.
The PNP-Conv blocks analyze the input with a dual-path convolution layer using a snake function~\cite{ziyin2020neural}, which is sensitive to periodic representations.
The high-level submodule employs periodic convolution (P-Conv) blocks to further extract the periodic components, while its following long short-term memory (LSTM) layer enables sequential modeling of the extracted periodic features.
In the estimation stage, we utilize a modified bi-directional feature pyramid network (BiFPN) to aggregate the multi-level features extracted in the analysis stage.
The multi-level feature fusion provides an accurate and effective pitch estimation by referencing various latent representations from each layer.
The designed neural network is optimized to pitch tracking task with only 0.362M parameters, indicating the effectiveness of the proposed composite modules and their ability to perform well without relying on high computational power or a large number of model parameters.
Our experiments on various datasets and ablation studies demonstrate the effectiveness of \shortname~compared to that of the baselines and highlight the importance of the submodules in extracting pitch components in various environments. 
In addition, our lightweight model, \shortname-S, exhibits notable pitch estimation accuracy despite having only 0.213M parameters which is equivalent to 59\% of the smallest baseline model.

\vspace{-5pt}

\section{Related works}

\subsection{Snake function}
In~\cite{ziyin2020neural}, the authors investigated the extrapolation properties of activation functions and proposed an effective activation function sensitive to periodicity, the Snake function.  
Neural networks that use the Snake function demonstrate impressive results in solving periodic problems such as atmospheric prediction, body temperature prediction, and financial data prediction. 
In addition, the Snake activation function was found to be advantageous in the optimization of model training compared to other periodic baselines.
The Snake function is formulated as follows:
\vspace{-3pt}
\begin{equation}
Snake_a(x)=x+\frac{1}{a}sin^2(ax),
\label{snake_eq}
\end{equation}
where $a$ denotes a pre-fixed constant that affects the frequency range the function focuses on. 
The authors reported that larger values of $a$ were more effective for input with periodic characteristics, while smaller values of $a$ were appropriate for non-periodic or standard tasks such as image classification.

\vspace{-3pt}

\subsection{BiFPN}
BiFPN~\cite{tan2020efficientdet} is one of the multi-scale fusion models~\cite{lin2017feature,ghiasi2019fpn} devised for object recognition.
BiFPN is a feature pyramid network that improves the target task with top-down and bottom-up pathways and cross-connections.
It matches the channel difference using convolution layers and the resolution difference by re-sampling.
Since different resolution features contribute unequally to the output, BiFPN provides a fast normalize fusion method with learnable weights.
With the lower network complexity and time cost, BiFPN achieve better object detection performance than that of the previous methods.

\section{Proposed model}
The speech signal exhibits a quasi-stationary characteristic, with its periodicity primarily arises from the pitch. 
Therefore, our proposed model, \shortname, is designed to be sensitive to the periodicity of speech signals, which is beneficial for accurate pitch estimation.
\shortname~emphasizes periodic characteristics in the latent representation and estimates pitch trajectory based on the representations.
The overall structure, as shown Figure~\ref{total_fig}, comprises two stages: analysis and estimation.
In the analysis stage, \shortname~extracts periodic information by eliminating non-periodic information in low-level representations.
The estimation stage tracks the pitch trajectory by leveraging the representation obtained from the analysis stage using a BiFPN module optimized for pitch estimation.

\vspace{-3pt}

\subsection{Analysis stage}

In the analysis stage, \shortname~analyzes speech signals by leveraging periodicity-sensitive modules, which include periodic convolution (P-Conv), non-periodic convolution (NP-Conv), and periodic and non-periodic convolution (PNP-Conv) blocks.
Our analysis structure is designed to first eliminate non-periodic information from the input and then enforce periodic characteristics.
In particular, the structure comprises two PNP-Conv blocks followed by three P-Conv blocks and an LSTM layer.
The PNP-Conv block consists of a dual-path convolution block, where one path is a P-Conv block, and the other is a NP-Conv block.
Both modules have two convolution layers in a stack, activated by the ReLU function and rectified by the Snake function.
The main difference between the P-Conv and NP-Conv block is the parameter $a$ in Eq.~\eqref{snake_eq}, which controls the frequency range of the periodicity.
Referring the findings in~\cite{ziyin2020neural}, the Snake function effectively processes periodic information with large $a$ values (5-50), while small $a$ (0.2-0.5) is suitable for processing non-periodic characteristics.
Therefore, we set $a$ as 0.2 for all NP-Conv blocks to eliminate non-periodic components, while $a$ of P-Conv blocks are set differently with larger values.
As the receptive field size of each layer increases, it becomes capable of capturing a larger range of temporal information.
Based on our preliminary experiments, we gradually reduced the values of $a$ in higher-level P-Conv blocks that have larger receptive fields, to (17, 13, 11, 7, 5) in order to increase sensitivity to the low-frequency range.
Subsequently, an LSTM layer enables powerful sequential modeling for pitch tracking.

\vspace{-3pt}

\subsection{Estimation stage}

In this work, we transform the pitch estimation problem into a classification task similar to~\cite{wei2022harmof0}, which estimates the level at which the pitch exists among 360 quantized levels of a limited frequency range.
Therefore, in the estimation stage, \shortname~estimates the discrete pitch frequency quantized in logarithmic scale by aggregating the multi-level features from the analysis stage using a modified BiFPN module.
In Figure~\ref{total_fig}, there is an overall structure of the BiFPN optimized for \shortname, which has half number of channels compared to the vanilla BiFPN.
The pre-sizing block aims to adjust the temporal resolution of the multi-level features to half that of the third-level feature ($P_3$), while maintaining the channel size. This is achieved through up- and down-sampling and using a depthwise separable convolution layer.
The resized five multi-level features are fused as below:
\vspace{-3pt}
\begin{equation} 
    P_i^{mid}=DSC\left({\frac{w_1\cdot P_i^{in}+w_2\cdot P_{i+1}^{in}}{w_1+w_2+\epsilon}}\right),
    \label{BiFPN_eq1}
\end{equation}
\begin{equation}
    P_i^{out}=DSC\left({\frac{w'_1\cdot P_i^{in}+w'_2\cdot P_i^{mid}+w'_3\cdot P_{i-1}^{out}}{w'_1+w'_2+w'_3+\epsilon}}\right),
    \label{BiFPN_eq2}
\end{equation}
where $P_i$ and $w_i$ denote the $i$-th level feature and its learnable weight factor, respectively. $DSC$ indicates the depthwise separable convolution layer activated by a Swish function~\cite{elfwing2018sigmoid}. The $\epsilon$ is set as 1e-4.

The BiFPN output is projected onto 360-dimensional quantized frequency bins using a projection layer, which has a fully-connected layer followed by a sigmoid function.
There is a 25-cent interval between consecutive quantization levels, and the frequency range is from 32.7Hz to 5834.5Hz.
The level can be converted to the frequency in Hertz using 
$f(i)=32.7\times 2^{25i/1200} ~[Hz]$,
where $i$ denotes the index of the level.

\subsection{Training criteria} \label{training_criteria}
The entire model is trained in an end-to-end manner, and we follow a similar training criterion as in~\cite{wei2022harmof0}, which involves minimizing the binary cross-entropy loss between the target one-hot vector $y$ and the predicted output vector $\hat{y}$ as follows:
\begin{equation}
    \mathcal{L}(y,\hat{y})=\sum_{i=1}^{360}\left(-y_i\log{\hat{y_i}}-(1-y_i)\log{(1-\hat{y_i})}\right)  
    \label{BCEloss_eq}
\end{equation}

\section{Experiments}

\subsection{Experimental details}\label{experimental_setting}

\noindent \textbf{Datasets.}
We trained and evaluated pitch estimation models in four different datasets that were resampled to 16kHz as listed.

\begin{itemize}
\item \textbf{VCTK-corpus (VCTK)}~\cite{veaux2017cstr} contains 44 hours of clean speech obtained from 109 speakers.
We used 100 speakers (40,212 utterances) for the training set and unseen 9 speakers (4,030 utterances) for the test set. 
The ratio of male to female numbers is close to one.

\item \textbf{PTDB-TUG (PTDB)}~\cite{pirker2011pitch}, typically used to evaluate the pitch tracking performance, consists of 576 minutes (4,720 utterances) of speech recorded by 20 English speakers. 
It includes laryngograph and reference pitch trajectories.

\item \textbf{MDB-stem-synth (MDB)}~\cite{salamon2017analysis} contains 418 minutes of 230 solo tracks, re-synthesized from the MedleyDB dataset~\cite{bittner2014medleydb}.
It consists of various instrumental sounds and singing voices with the F0 annotations.

\item \textbf{MIR-1k (MIR)}~\cite{hsu2009improvement} contains 133 minutes of singing voices (11 males, 8 females) recorded with the musical accompaniment, and pitch annotations.
\end{itemize}

\noindent We splitted the datasets into training, validation, and test sets in ratio of 3:1:1, except for the VCTK dataset.
To evaluate the robustness of models to environmental distortion, we created a dataset called VCTK-Distortion (VCTK-DT).
This dataset was generated by convolving speech signals from the VCTK dataset with room impulse responses obtained from the MIT Impulse Response Survey~\cite{traer2016statistics}, and adding noise from the NOISEX-92 dataset~\cite{varga1993assessment}.
The signal-to-noise ratio (SNR) was randomly selected from the ranges (-7, -2, 3, 8, 13) dB for the training set and uniformly selected from the ranges (-5, 0, 5, 10, 15) dB for the test set.
To acquire ground-truth pitch trajectory, we used DIO~\cite{morise2009fast, morise2010rapid} algorithm.

\begin{figure*}[t]
   \footnotesize
    \begin{minipage}[t]{0.26\linewidth}
        \centering
        \begin{tikzpicture}
        \pgfplotsset{compat=1.4}
        \begin{axis}[ 
        width=1.3\columnwidth, height=\linewidth,
        scaled y ticks = false,
        scaled x ticks = false,
        xtick pos=left, ytick pos=left,
        ybar, ymode=log, 
        bar width=2pt,
        xmin=-.5, xmax=5.5,
        ymin=0.7,   ymax=21,
        xlabel={SNR (dB)}, ylabel={MAE (Hz)},
        xticklabels={-5, 0, 5, 10, 15, clean},
        xtick={0, 1, 2, 3, 4, 5},
        ytick= {0, 0.5, 1, 2.5, 5, 10, 20},
        yticklabels={0, 1, 2.5, 5, 10, 20,},
        minor tick length=1ex,
        major x tick style = {opacity=1},
        ymajorgrids=true,
        xmajorgrids=false,
        grid style=dashed,
        legend style={at={(0.7497, 1.8)},
        anchor=north, 
        legend columns=-3
        },
        ]
        \addplot[black, fill=gray] coordinates{
        (0, 12.11) (1,9.279) (2,7.94) (3,6.187) (4,5.829) (5,3.33) }; 
        \addplot[black, fill=turquoise] coordinates{
        (0, 18.12) (1,12.61) (2,9.24) (3,6.687) (4,5.96) (5,3.249)}; 
        \addplot[black, fill=orange] coordinates{
        (0, 3.154) (1,2.12) (2,1.667) (3,1.312) (4,1.309) (5,0.8224)}; 
        \addplot[black, fill=darkred] coordinates{
        (0, 2.697) (1,1.734) (2,1.441) (3,1.148) (4,1.188) (5,0.7217)}; 
        \end{axis}
        \end{tikzpicture}
        \centerline{\qquad\qquad\qquad(a)}
    \end{minipage}
    \hfill
    \begin{minipage}[t]{0.26\linewidth}
    \centering
        \begin{tikzpicture}
        \pgfplotsset{compat=1.4}
        \begin{axis}[
        width=1.3\columnwidth, height=\linewidth,
        scaled y ticks = false,
        scaled x ticks = false,
        xtick pos=left, ytick pos=left,
        ybar,
        bar width=2pt,
        enlargelimits=0.005,
        xmin=-.5, xmax=5.5,
        ymin=50,   ymax=98,
        xlabel={SNR (dB)}, ylabel={RPA (\%)},
        xticklabels={-5, 0, 5, 10, 15, clean},
        xtick={0, 1, 2, 3, 4, 5},
        ytick={50, 55, 60, 65, 70, 75, 80, 85, 90, 95},
        yticklabels={, 55, , 65, , 75, , 85, , 95},
        minor tick length=1ex,
        major x tick style = {opacity=1},
        ymajorgrids=true,
        xmajorgrids=false,
        grid style=dashed,
        legend style={at={(0.7497, 1.8)},
        anchor=north, legend columns=-3},
        ]
        \addplot[black, fill=gray] coordinates{
        (0,64.13) (1,71.61) (2,75.34) (3,79.33) (4,81.78) (5,89.92) };  
        \addplot[black, fill=turquoise] coordinates{
        (0, 50.52) (1,63.87) (2,71.47) (3,76.85) (4,80.12) (5,90.82) }; 
        \addplot[black, fill=orange] coordinates{
        (0, 81.88) (1,86.19) (2,88.55) (3,89.83) (4,91.04) (5,95) }; 
        \addplot[black, fill=darkred] coordinates{
        (0, 83.97) (1,88.34) (2,90.56) (3,91.7) (4,92.87) (5,96.48) }; 
        \end{axis}
        \end{tikzpicture}
        \centerline{\qquad\qquad\qquad(b)}
    \end{minipage}
    \hfill
    \begin{minipage}[t]{0.26\linewidth}
    \centering
        \begin{tikzpicture}
        \pgfplotsset{compat=1.4}
        \begin{axis}[
        width=1.3\columnwidth, height=\linewidth,
        scaled y ticks = false,
        scaled x ticks = false,
        xtick pos=left, ytick pos=left,
        ybar,
        bar width=2pt,
        enlargelimits=0.005,
        xmin=-.5, xmax=5.5,
        ymin=50,   ymax=98,
        xlabel={SNR (dB)}, ylabel={RCA (\%)}, 
        xticklabels={-5, 0, 5, 10, 15, clean},
        xtick={0, 1, 2, 3, 4, 5},
        ytick={50, 55, 60, 65, 70, 75, 80, 85, 90, 95},
        yticklabels={, 55, , 65, , 75, , 85, , 95},
        minor tick length=1ex,
        major x tick style = {opacity=1},
        ymajorgrids=true,
        xmajorgrids=false,
        grid style=dashed,
        legend style={at={(0.7497, 1.8)},
        anchor=north, legend columns=-3},
        ]
        \addplot[black, fill=gray] coordinates{ 
        (0,68.76) (1,75.12) (2,78.66) (3,82.25) (4,84.06) (5,91.23) };  
        \addplot[black, fill=turquoise] coordinates{
        (0, 54.97) (1,66.26) (2,73.62) (3,78.52) (4,81.1) (5,91.33) }; 
        \addplot[black, fill=orange] coordinates{
        (0, 81.89) (1,86.19) (2,88.55) (3,89.83) (4,91.04) (5,95.01) }; 
        \addplot[black, fill=darkred] coordinates{
        (0, 83.97) (1,88.34) (2,90.56) (3,91.71) (4,92.88) (5,96.48) }; 
        \end{axis}
        \end{tikzpicture}
        \centerline{\qquad\qquad\qquad(c)}
    \end{minipage}
    \hfill
    \vspace{-5pt}
    \caption{Pitch estimation performances in various SNRs on the VCTK-DT test set. (a) MAE (Hz) in log-scale; (b) RPA (\%); (c) RCA (\%). The \tgray{gray}, \tblue{blue}, \torange{yellow}, and \tred{red} bars indicate the results for CREPE, DeepF0, HarmoF0, and the proposed model \shortname. The low MAE, high RPA, and RCA indicate better performance.}
    \label{fig:SNR_analysis_graph}
    \vspace{-7pt}
\end{figure*}
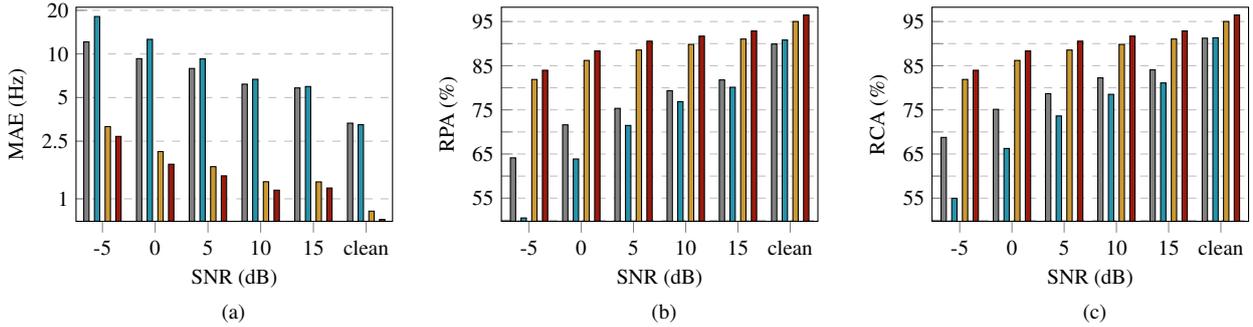

\begin{table}[]
\centering
\caption{Performance results on four clean datasets. Average raw pitch accuracy (RPA) and raw chroma accuracy (RCA). Both high RPA and RCA scores indicate better performances.}
\vspace{-5pt}
\label{tab:baselines}
\resizebox{\columnwidth}{!}
{
\begin{tabular}{l||c|c|cccc}
\toprule
\textbf{Model}                     & \begin{tabular}[c]{@{}c@{}}\textbf{Params.}\\ \textbf{(M)}\end{tabular} & \begin{tabular}[c]{@{}c@{}}\textbf{Metrics}\\ \textbf{($\%$)}\end{tabular} & \textbf{VCTK}  & \textbf{PTDB}  & \textbf{MDB}   & \textbf{MIR}   \\ \midrule
\multirow{2}{*}{pYIN~\cite{mauch2014pyin}}     & \multirow{2}{*}{$-$}                                    & RPA $\uparrow$                                                   &  54.20     & 50.51 & 90.12 & 90.47 \\
                          &                                                       & RCA$\uparrow$                                                    & 55.00      & 51.30 & 90.71 & 91.06 \\ \midrule
\multirow{2}{*}{SWIPE~\cite{camacho2008sawtooth}}    & \multirow{2}{*}{$-$}                                    & RPA $\uparrow$                                                    & 77.74   & 67.45 & 92.50 & 96.36 \\
                          &                                                       & RCA$\uparrow$                                                    & 73.44      & 69.50 & 93.34 & 96.73 \\ \midrule
\multirow{2}{*}{CREPE~\cite{kim2018crepe}}    & \multirow{2}{*}{22.240}                                 & RPA $\uparrow$                                                    & 89.92 & 81.44 & 96.34 & 96.41 \\
                          &                                                       & RCA$\uparrow$                                                 & 91.23 & 84.26 & 96.74 & 96.72 \\ \midrule
\multirow{2}{*}{DeepF0~\cite{singh2021deepf0}}   & \multirow{2}{*}{4.961}                                    & RPA $\uparrow$                                                   & 90.82 & 93.14 & 98.38 & 97.82 \\
                          &                                                       & RCA$\uparrow$                                                    & 91.33 & 93.47 & 98.44 & 98.28 \\ \midrule
\multirow{2}{*}{HarmoF0~\cite{wei2022harmof0}}  & \multirow{2}{*}{0.377}                                & RPA $\uparrow$                                                   & 95.00 & 93.56 & 98.40 & 98.34 \\
                          &                                                       & RCA$\uparrow$                                                    & 95.01 & 93.59 & 98.46 & 98.46 \\ \midrule
\multirow{2}{*}{\bf MF-PAM}   & \multirow{2}{*}{0.362}                                & RPA $\uparrow$                                                   & \bf 96.62 & \bf 97.12 & \bf 99.20 & \bf 98.97 \\
                          &                                                       & RCA$\uparrow$                                                    & \bf 96.62 & \bf 97.13 & \bf 99.20 & \bf 98.99 \\ \midrule
\multirow{2}{*}{\bf MF-PAM-S} & \multirow{2}{*}{\bf 0.213}                                & RPA $\uparrow$                                                   & 96.33 & 96.62 & 99.05 & 98.93 \\
                          &                                                       & RCA$\uparrow$                                                    & 96.33 & 96.62 & 99.05 & 98.96 \\ \bottomrule
\end{tabular}

}

\vspace{-10pt}
\end{table}

\vspace{1pt}
\noindent \textbf{Network configurations.}
The input ($C_{in}$) and output ($C_{out}$) channel sizes of the PNP-Conv and P-Conv blocks are (1, 6, 12, 24, 48) and (6, 12, 24, 48, 96), respectively, with a stride of 4 and dilation of 1.
The kernel sizes ($K$) are sequentially increased by 4, 4, 8, 8, and 12 in each block to utilize a larger receptive field.
For the light BiFPN, depthwise separable convolution layers have a kernel size of 5, strided and dilated by 1.
In addition, we up-sampled the input by a factor of 4 using the sinc interpolation filter~\cite{smith1984flexible} before the analysis stage to provide richer context information.

\vspace{1pt}

\noindent \textbf{Evaluation protocols.}
We evaluated the pitch estimation performance using raw pitch accuracy (RPA) and raw chroma accuracy (RCA)~\cite{salamon2014melody}, and the threshold was set to 50 cents.
RPA and RCA measure the percentage of the number of frames, in which the pitch errors are smaller than the threshold value.
The difference between RCA and RPA is that the RCA ignores the error by a single octave since the chroma represents 12 different pitch classes without the concept of an octave in musical datasets.
We measured the mean absolute error (MAE) on the VCTK dataset to evaluate the pitch error in Hz.

\vspace{-3pt}
\subsection{Results}

To evaluate the accuracy of the estimated pitch, we compared the pitch estimation performance of the proposed model with the two signal processing based methods (pYIN~\cite{mauch2014pyin}, SWIPE~\cite{camacho2008sawtooth}) and three deep learning-based models (CREPE~\cite{kim2018crepe}, DeepF0~\cite{singh2021deepf0}, HarmoF0~\cite{wei2022harmof0}).
\begin{table}[]
\centering
\caption{Performance results on VCTK-DT. Average mean absolute error (MAE), raw pitch accuracy (RPA), and raw chroma accuracy (RCA). Up arrow indicates that higher score is better while down arrow indicates lower score is better.}
\vspace{-5pt}
\label{tab:various_env}
\resizebox{\columnwidth}{!}{%
\begin{tabular}{l||c|c|rrr|r}
\toprule
{\bf Model}                     & \begin{tabular}[c]{@{}c@{}}\textbf{FLOPs\quad}\\ \textbf{(G)\quad}\end{tabular} & {\bf Metrics}  & {\bf All}                       & {\bf Noise}                     & {\bf Reverb}                    & {\bf Clean}                     \\ \midrule
\multirow{3}{*}{CREPE~\cite{kim2018crepe}}    & \multirow{3}{*}{480.875} & MAE (Hz)$\downarrow$ & 8.24                     & 6.59                     & 5.12                     & 3.33                     \\
                                                                        && RPA~~~(\%) $\uparrow$ & 74.51                     & 80.43                     & 83.31                     & 89.92                     \\
                                                                        && RCA~~(\%) $\uparrow$ & 77.83                     & 83.15                     & 85.36                     & 91.23                     \\ \midrule
\multirow{3}{*}{DeepF0~\cite{singh2021deepf0}}   &\multirow{3}{*}{1378.921}& MAE (Hz)$\downarrow$ & 10.49                     & 8.24                     & 5.54                     & 3.25                     \\
                          && RPA~~~(\%) $\uparrow$ & 68.66                     & 77.82                     & 80.56                     & 90.82                     \\
                          && RCA~~(\%) $\uparrow$ & 70.97                     & 79.28                     & 81.68                     & 91.33                     \\ \midrule
\multirow{3}{*}{HarmoF0~\cite{wei2022harmof0}}  &\multirow{3}{*}{43.705}& MAE (Hz)$\downarrow$ & 1.91                     & 1.66                     & 1.22                     & 0.82 \\                     
                          && RPA~~~(\%) $\uparrow$ & 87.52                     & 90.52                     & 91.18                     & 95.00                     \\
                          && RCA~~(\%) $\uparrow$ & 87.53                     & 90.53                     & 91.19                     & 95.01                     \\ \midrule
\multirow{3}{*}{\bf MF-PAM}   &\multirow{3}{*}{\textbf{0.101}} & MAE (Hz)$\downarrow$ & \bf 1.64                     & \bf 1.35                     & \bf 1.10                     & \bf 0.69 \\      
                          && RPA~~~(\%) $\uparrow$ & \bf 90.05                     & \bf 92.20                     & \bf 93.29                     & \bf 96.62                     \\
                          && RCA~~(\%) $\uparrow$ & \bf 90.05                     & \bf 92.20                     & \bf 93.29                     & \bf 96.62                     \\ \midrule
\multirow{3}{*}{\bf MF-PAM-S} & \multirow{3}{*}{\textbf{0.101}} & MAE (Hz)$\downarrow$ & 2.08                     & 1.56                     & 1.30                     & 0.88 \\                    
                          && RPA~~~(\%)  $\uparrow$ & 87.89                     &91.82                       &91.45                     &95.32                     \\
                          && RCA~~(\%) $\uparrow$ & 87.89 & 91.82 & 91.45 & 95.32 \\ \bottomrule
\end{tabular}%
}
\vspace{-10pt}
\end{table}

\vspace{1pt}
\noindent \textbf{Comparison with baseline models.}
Table \ref{tab:baselines} shows the pitch estimation performance of the proposed model and baselines on the four clean datasets.
Our proposed model outperformed baselines in every metric and dataset.
In general, the number of periodic components in VCTK and PTDB datasets is less than that of musical datasets.
Thus the pitch estimation performance of the baseline models showed a more severe degradation than that of \shortname.
For the MDB dataset, our model showed an estimation accuracy of $>$99\% in terms of RCA and RPA, even with the smallest number of network parameters (0.362 M).
We achieved higher pitch estimation performance with over 40 \% fewer parameters by eliminating the LSTM layer in \shortname~(\shortname-S), compared to HarmoF0.
These results indicate the effectiveness of the proposed modules.

Figure~\ref{fig:SNR_analysis_graph} depicts the estimation performance of the baseline models and \shortname~in various SNRs based on the VCTK-DT.
Evidently, \shortname~significantly outperforms the baselines across all the metrics (MAE, RPA, and RCA), especially in -5 and 0 dB SNRs.
Table \ref{tab:various_env} presents the model performance in various environments; Clean, Noise, Reverberation (Reverb), and all distortions (All) based on the VCTK-DT.
The table demonstrates that \shortname~accurately estimated the pitch in all environments and exhibited a minor performance degradation in harsh conditions.

\begin{figure}[t]
   \footnotesize
   \centering
    \begin{minipage}[t]{0.26\linewidth}
        \centering
        \begin{tikzpicture}
        \pgfplotsset{compat=1.3}
        \begin{axis}[ 
        width=3.8\columnwidth, height=2.8\linewidth,
        scaled y ticks = true,
        scaled x ticks = true,
        xtick pos=left, ytick pos=left,
        xmin=-.5, xmax=5.5,
        ymin=79,   ymax=97,
        xlabel={SNR (dB)}, ylabel={RPA (\%)},
        xticklabels={-5, 0, 5, 10, 15, clean},
        xtick={0, 1, 2, 3, 4, 5},
        ytick= {79, 81, 83, 85, 87, 89, 91, 93, 95, 97 },
        yticklabels={79, 81, 83, 85, 87, 89, 91, 93, 95, 97},
        minor tick length=1ex,
        major x tick style = {opacity=1},
        ymajorgrids=true,
        xmajorgrids=true,
        grid style=dashed,
        legend style={at={(0.5, 1)},
        anchor=west, 
        legend columns=1
        },
        legend pos=south east,
        legend cell align={left}
        ]
        \addplot[ mark=square,darkred] plot coordinates{
        (0, 84.02) (1,88.43) (2,91.14) (3,92.01) (4,93.01) (5,96.62) }; 
        \addplot[ mark=triangle, turquoise] plot coordinates{
        (0, 81.22) (1,86.67) (2,90.1) (3,91.56) (4,92.87) (5,96.61)}; 
        \addplot[ mark=diamond, purple] plot coordinates{
        (0, 82.02) (1,87.46) (2,90.31) (3,91.43) (4,92.64) (5,96.48)}; 
        \addplot[ mark=diamond, gray] plot coordinates{
        (0, 82.32) (1,87.28) (2,89.9) (3,91.58) (4,92.81) (5,96.67)}; 
        \addplot[ mark=o, orange] plot coordinates{
        (0, 80.04) (1,85.86) (2,88.62) (3,90.57) (4,92.25) (5,95.87)}; 
        \scriptsize{
        \addlegendentry{\textbf{MF-PAM (Proposed)}}
        \addlegendentry{w/o PNP-Conv block}
        \addlegendentry{w/o Snake}}
        \addlegendentry{w/o multi-level}
        \addlegendentry{w/o Light BiFPN}

        \end{axis}
        \end{tikzpicture}
    \end{minipage}
    \hfill
    \vspace{-5pt}
    \caption{Ablation study for each module in various SNR levels on the VCTK-DT test set. Average raw pitch accuracy (RPA). High RPA scores indicate better performance.}
    \label{fig:ablation_study}
    \vspace{-7pt}
\end{figure}
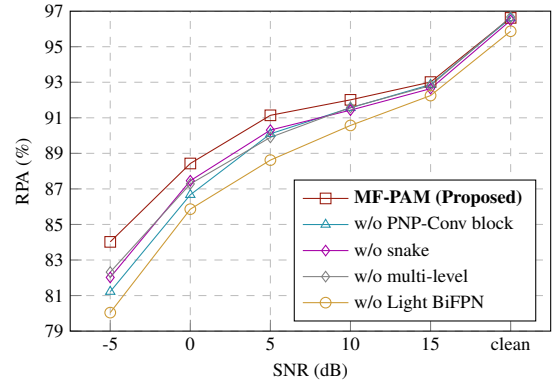

\vspace{1pt}
\noindent \textbf{Ablation study.}
We further investigated the individual contributions of each proposed module, the PNP-Conv block and the light BiFPN as shown in Figure~\ref{fig:ablation_study}.
We replaced the PNP-Conv block with the P-Conv block with a larger hidden channel size to match the model size with the `w/o PNP-Conv block' setup.
For the `w/o light BiFPN' setup, we removed the light BiFPN layer, and for the `w/o multi-level' setup, we only used the last feature of the analysis stage as the input for the light BiFPN.
As shown in Figure~\ref{fig:ablation_study}, while the pitch estimation accuracy of `w/o PNP-Conv' was similar to that of \shortname~in the clean environment (RPA: 96.62\% vs. 96.07\%), \shortname~demonstrated more robust pitch estimation performance in low SNRs compared to `w/o PNP-Conv' (RPA: 84.52\% vs. 81.28\% in -5 dB SNR).
The results demonstrate that the PNP-Conv block encouraged the proposed model to extract only pitch-related information even in extremely noisy conditions.
Although `w/o multi-level' has a bigger model size than `w/o light BiFPN', both showed similar pitch estimation performance (RPA: 95.42\% vs. 95.32\% in clean signal).
These results indicate that the light BiFPN architecture as well as the multi-level features are crucial for improving the pitch estimation performance.


\section{Conclusions}

In this paper, we proposed a novel pitch estimation model, \shortname\footnote{\url{https://github.com/Woo-jin-Chung/MF-PAM_mfpam_pitch_estimation_pytorch}}, which extract periodic-related information effectively from the raw audio input using periodicity-sensitive blocks.
The pitch-related representation was processed by leveraging a multi-level feature fusion model, BiFPN, and projected onto quantized frequency levels for the pitch estimation.
Our experimental results demonstrated that \shortname~outperformed state-of-the-art baseline models in various datasets and conditions, thanks to its structural configurations that consider the periodicity of speech signals.
We further conducted ablation studies to investigate the contributions of the submodules of \shortname and confirmed their effectiveness.
Moreover, the lightweight version of our proposed model, \shortname-S, achieved competitive performance in terms of RPA and RCA with significantly fewer parameters, over 40\% less than the smallest baseline model.
%


\clearpage
\bibliographystyle{IEEEtran}
\bibliography{longstrings,mybib}
\clearpage
\end{document}